\documentclass[prl,preprint,showpacs]{revtex4} %showpacs
\usepackage{epsfig}
\usepackage{amsmath}

\begin{document}

\title{Spin injection in the non-linear regime: band bending effects}

\author{G. Schmidt, C. Gould, P. Grabs, A.M. Lunde,\cite{byline} G. Richter, A. Slobodskyy,
and L.W. Molenkamp}

\affiliation{Physikalisches Institut (EP3), Universit\"at
W\"urzburg, Am Hubland, 97074 W\"urzburg, Germany}

\date{\today}

\begin{abstract}

Semiconductor spintronics will need to control spin injection
phenomena in the non-linear regime. In order to study these
effects we have performed spin injection measurements from a
dilute magnetic semiconductor [(Zn,Be,Mn)Se] into nonmagnetic
(Zn,Be)Se at elevated bias. When the applied voltage is increased
to a few mV we find a strong decrease of the spin injection
efficiency. The observed behavior is modelled by extending the
charge-imbalance model for spin injection to include band bending
and charge accumulation at the interface of the two compounds. We
find that the observed effects can be attributed to repopulation
of the minority spin level in the magnetic semiconductor.
\end{abstract}


\maketitle

Much of our understanding of the physics of electrical spin
injection from a ferromagnet into a normal metal derives from a
current-imbalance model proposed by van Son\cite{vanson} and,
independently, by Johnson and Silsbee\cite{johnsonsilsbee}. In
this model, the current conversion is driven by a splitting of the
Fermi levels for majority- and minority-spin electrons at the
ferromagnet-normal metal interface. The concomitant reduction of
the conductivity of the normal metal due to the suppression of a
spin-channel leads to a discontinuity of the average
electrochemical potential, which is sometimes referred to as the
spin-induced boundary resistance. The model was put on a solid
(Boltzmann) basis by Valet and Fert\cite{valetfert} and proved
very useful for describing important spin transport phenomena
including CPP-GMR (giant magnetoresistance in the
current-perpendicular-to-plane geometry)\cite{cpp}.

In a further application, we used the model to describe spin
injection into semiconductors\cite{gs}, revealing the importance
of a conductance mismatch between ferromagnetic metals and
semiconductor. The conductance mismatch keeps the splitting of the
spin-dependent Fermi levels at minimum, thus precluding spin
injection. In the meantime, several methods of avoiding the
conductivity mismatch have been proposed. However, by far the most
robust route towards spin injection to date\cite{fiederling,LMR}
is the use of dilute II-VI magnetic semiconductors (DMSs) that
exhibit the giant Zeeman effect\cite{dms}, have a conductivity
comparable to that of non-magnetic semiconductors, and can boast
spin polarizations close to 100 \%  at a small applied magnetic
field.

Recently, we used a simple DMS - non-magnetic semiconductor (NMS)
heterostructure (consisting of (Zn,Mn,Be)Se as DMS and lattice
matched (Zn,Be)Se as NMS) to demonstrate\cite{LMR} the field
dependence of the spin-induced boundary resistance, which
increases with field as the magnetization of the paramagnetic DMS
increases. The experiments reported in Ref. \cite{LMR} were all
done in the regime of linear response, where the current-imbalance
model is appropriate.

However, spin injection experiments in semiconductors allow one to
very easily enter the regime of non-linear response, where
corrections to the current-imbalance model are necessary. These
corrections will be manifest in any future spintronic
semiconductor device, and it is thus of importance to identify
them correctly.

In this paper we report on spin-injection measurements in the
non-linear regime. More specifically, we experimentally
demonstrate the strong effects that band-bending and
charge-accumulation have on the non-linear transport. We present a
modelling of the observed phenomena which generalizes the
charge-accumulation model to include the limited screening
available in semiconductor structures, and shows good agreement
with the experiments. It should be noted that the effects
discussed here involve non-linearities influencing the spin
polarization at the DMS/NMS interface, and thus are of a
fundamentally different nature than the drift-induced effects
discussed recently by Yu and Flatt\'{e}\cite{flatte} (which occur
at higher electric fields).

The devices on which the experiments were carried out consisted of
an all II-VI semiconductor heterostructure fabricated by molecular
beam epitaxy. The heterostructure consisted of three semiconductor
layers. From bottom to top these layers are a non magnetic n-type
Zn$_{0.97}$Be$_{0.03}$Se layer (thickness 500 nm, $ n \approx
10^{19}$ cm$^{-3}$), a dilute magnetic
Zn$_{0.89}$Be$_{0.05}$Mn$_{0.06}$Se layer (thickness 100 nm, $n
\approx 5 \times 10^{18}$ cm$^{-3}$) which acts as a spin aligner,
and a top layer of 30 nm highly n-doped ZnSe ($ n \approx 2 \times
10^{19}$ cm$^{-3}$). The latter was grown to ensure good quality
ohmic contacts and was covered in-situ with aluminium without
breaking the UHV.

\begin{figure}
\centerline {\psfig{file=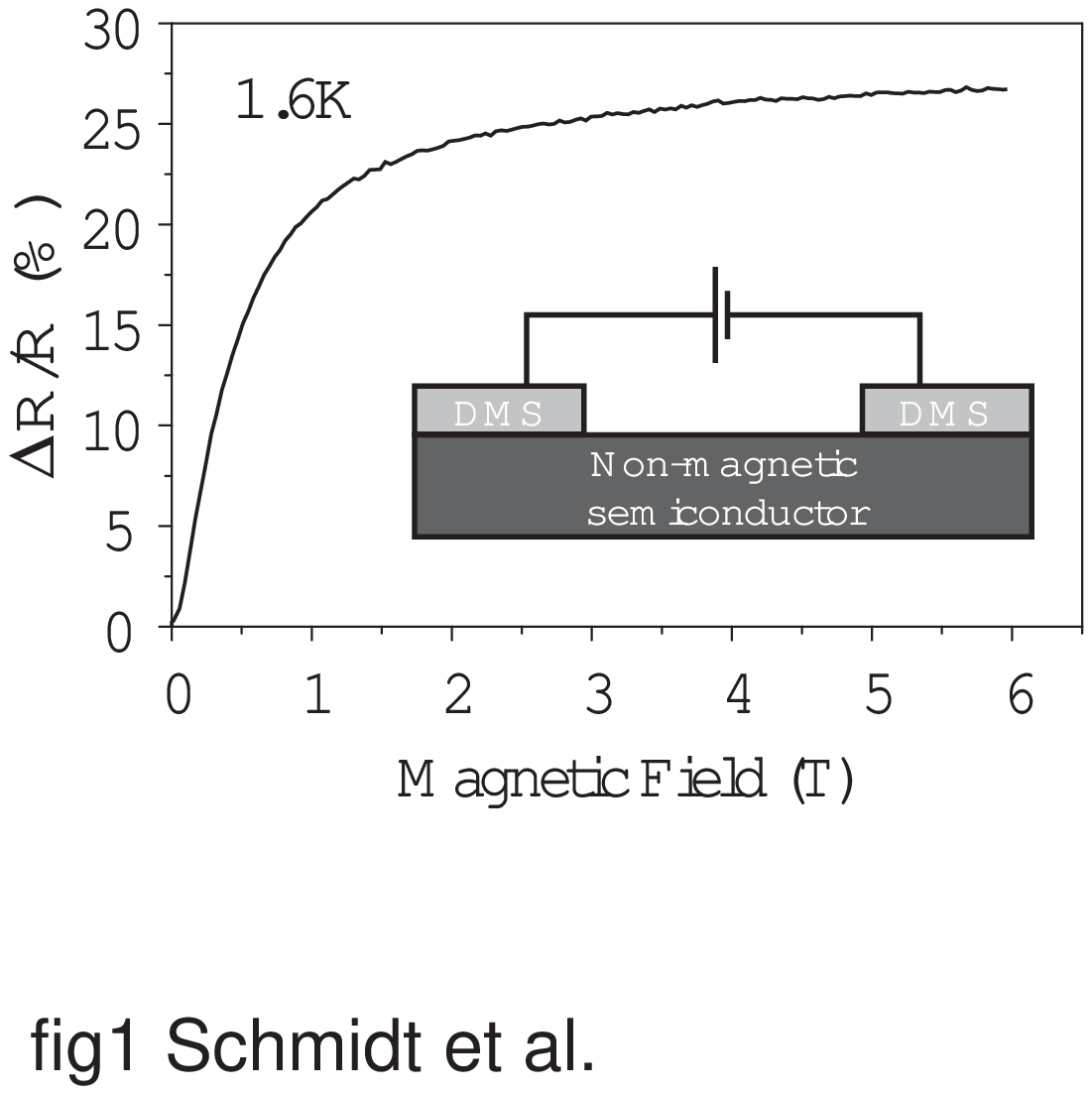,width=14.0cm}} \caption{Insert:
Spin injection device used in the experiment consisting of a
non-magnetic semiconductor layer with two DMS top-contacts. The
graph gives the resistance change $\Delta R / R$ versus magnetic
field $B$. } \label{fig1}
\end{figure}

In the Al-layer,  $200 \times 250 \; \mu$m contact pads were
defined by optical lithography and wet chemical etching. These
pads were used as a mask for a subsequent wet etching step through
the ZnSe and the DMS down to the Zn$_{0.97}$Be$_{0.03}$Se. A mesa
was then defined by etching down to the substrate, leaving only
two contacts and the transport layer in between. The resulting
sample is schematically shown as the inset in Fig. \ref{fig1}.

The samples were immersed in a magnetocryostat and their transport
properties were determined at temperatures of 1.6 K, 3 K, 4.2 K
and 6 K. The magnetoresistance of the devices was measured using
dc techniques and a quasi four probe geometry, in which the wiring
resistance of the setup was excluded, while the contact resistance
of the device was still part of the measured
resistance.\cite{4probe} For bias voltages $V_{\rm bias}$ in the
regime of linear response (300$\mu$V or less) the device showed a
positive magnetoresistance. Fig. \ref{fig1} plots the relative
magnetoresistance  $\Delta R / R$ for a sample with a distance
$x_0=$ 10 $\mu$m between the contact pads, taken at 1.6 K, where
the zero-field resistance $R$ = 420 $\Omega$. As described in Ref.
\cite{LMR}, the magnetoresistance results from the increase of the
spin-induced boundary resistance with magnetic field at the two
DMS/NMS interfaces in the samples. All data discussed here were
taken on the same sample as that for which the low-temperature
magnetoresistance is shown in Fig. \ref{fig1}; we verified that
the effects discussed occur in samples with varying doping
concentrations and dimensions. We found experimentally that both
$R$ and the saturated magnetoresistance $\Delta R /R \approx$ 0.25
are independent of temperature in the investigated range.

\begin{figure}
\centerline {\psfig{file=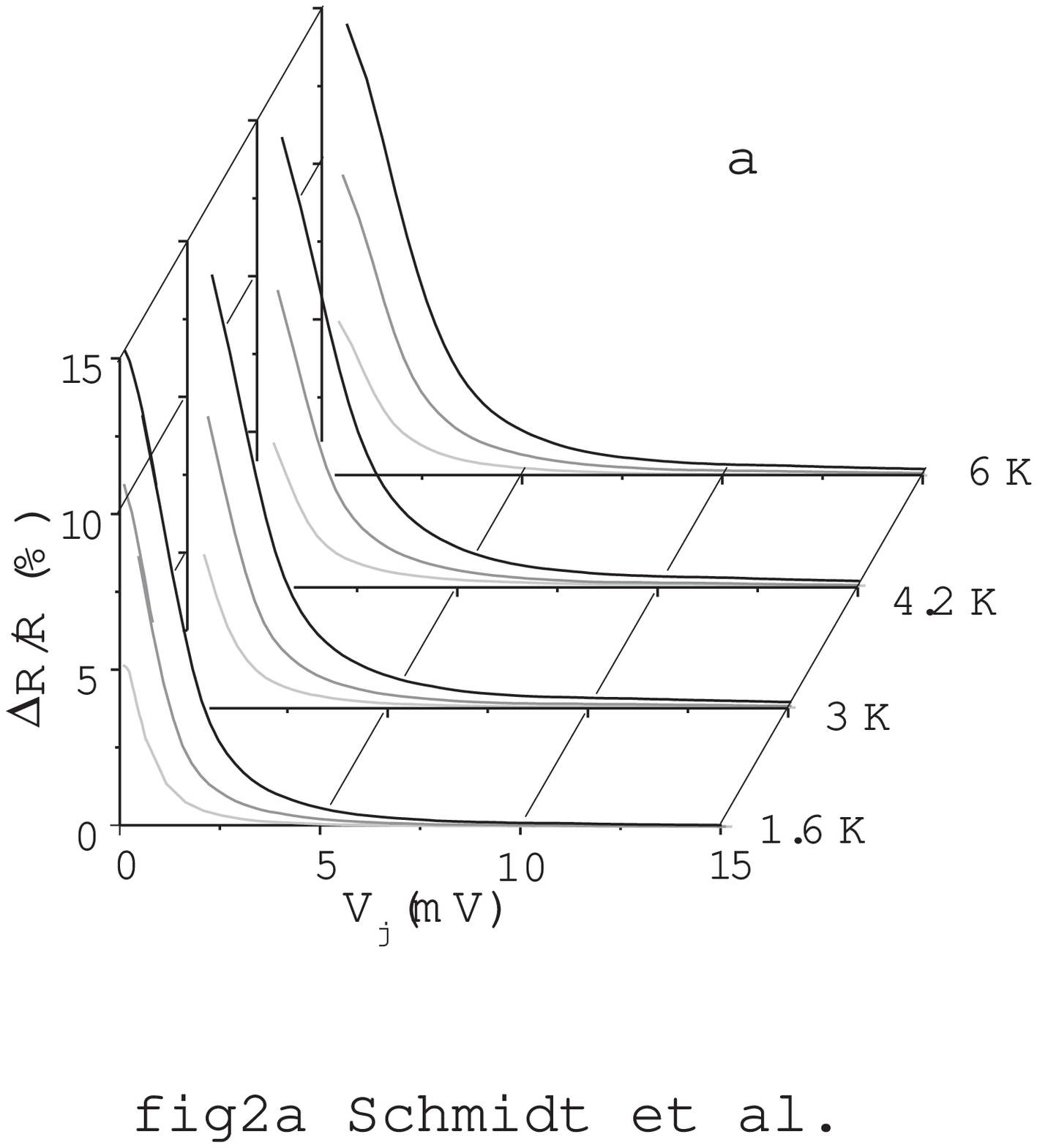,width=8.0cm}} \centerline
{\psfig{file=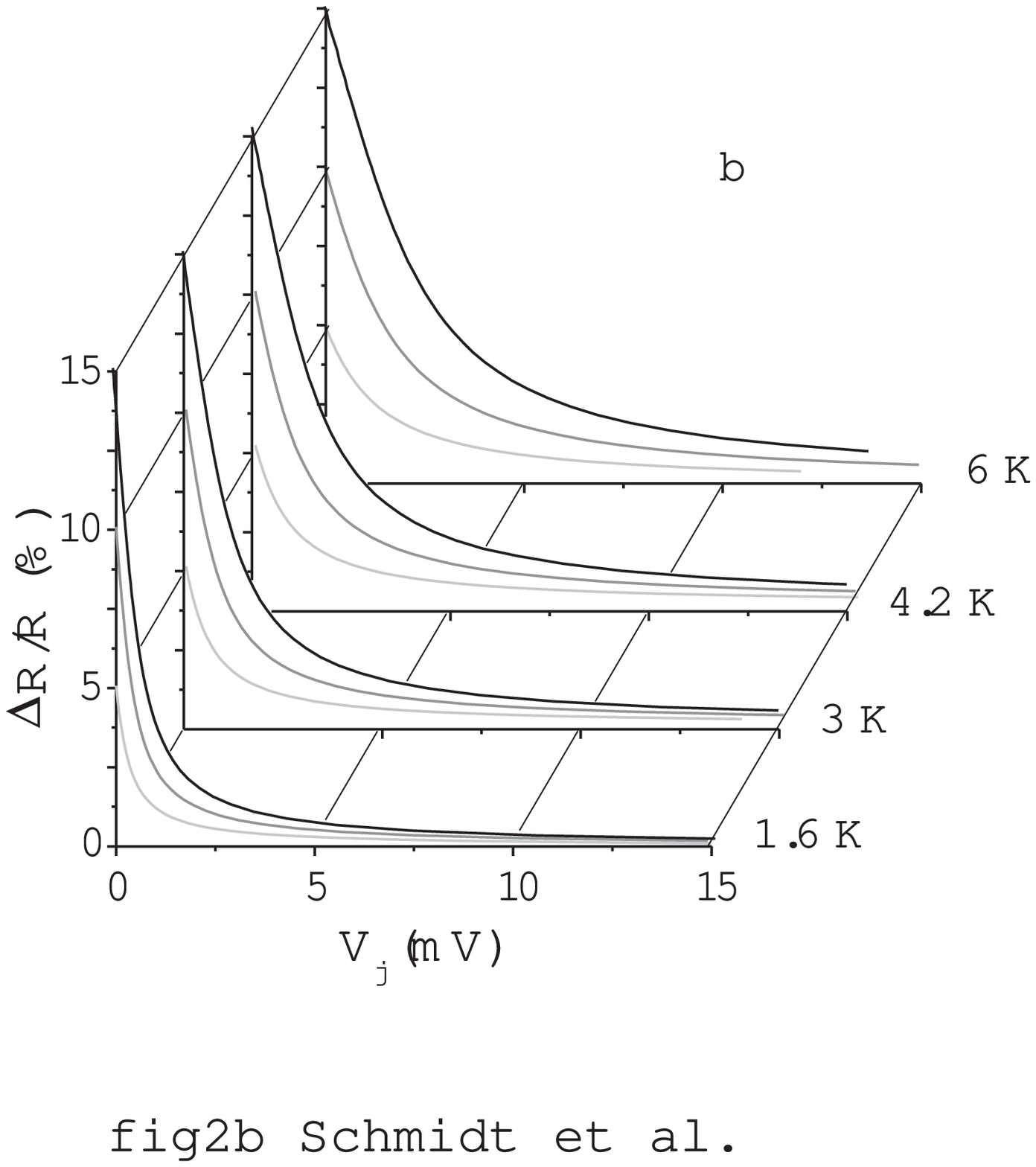,width=8.0cm}} \caption{(a) Experimental
and (b) theoretical non-linear magnetoresistance $\Delta R / R$
data plotted over voltage drop $V_j$. To facilitate comparison
between experiment and theory, curves are plotted starting at
several fixed values of $\Delta R / R $ (obtained by carefully
adjusting $B$), for temperatures of 1.6, 3, 4.2 and 6 K. The
parameters involved in the modelling are discussed in the text.
}\label{exp}
\end{figure}

The main experimental result of this paper is shown in
Fig.\ref{exp}(a). When the applied voltage is increased, a
pronounced and very rapid drop of the magnetoresistance is
observed, reducing the effect by two or more orders of magnitude
on applying a voltage of around 10 mV across the junction. The
experimental data in Fig. 2a were taken starting from three
different values of $\Delta R / R$ (i.e., at different values of
the magnetic field $B$) in the linear response regime (i.e.,
$\Delta R / R \approx $ 0.05, 0.1 and 0.15, respectively), at the
four different temperatures mentioned above. Obviously, the
non-linearities show a marked temperature dependence. Moreover,
the horizontal axis displays the voltage drop over the junction
$V_j$, rather than the total bias $V_{\rm bias}$ (roughly, $V_j
\approx 0.15 V_{\rm bias} $), for reasons of comparison with the
modelling discussed below, and because it more clearly illustrates
the energy scales involved in the non-linearities. We will detail
below how $V_j$ is defined and can be calculated from $V_{\rm
bias}$.

The drop of the magnetoresistance can be understood if we combine
the model for diffusive spin polarized transport with the band
structure of the semiconductor heterostructure. When a current is
driven from a spin polarized material into a non polarized
material, the electrochemical potentials for spin up
($\mu^\uparrow$) and spin down ($\mu^\downarrow$) split at the
interface. In linear response,\cite{flatte2} the length scale of
this splitting is given by the spin scattering length of each
material. The situation is depicted in Fig. 3, where the
Zeeman-split conduction band (full drawn lines) and relevant
potentials (dash-dotted lines) of the DMS are shown of the left
side of the figure. The interface is indicated by the dotted
vertical line at $x=0$, with the NMS in the right half of the
plane. The splitting of ($\mu^\uparrow$) and ($\mu^\downarrow$) is
the driving force which leads to a spin polarized current in the
non magnetic material. Because the conductivities for spin up and
spin down are equal in the NMS, only a difference in the
derivative of the electrochemical potential can lead to different
currents in both spin channels. Since the {\it electrical}
potential must be equal for both spin directions, this difference
can only be introduced through the {\it chemical} potential, i.e.,
by spin accumulation. Spin injection thus leads to a potential
drop at the interface which drives the spin conversion. This
voltage drop, which may alternatively be regarded as a
spin-induced boundary resistance (since it represents an extra
voltage drop in the sample in absence of spin injection), is
indicated in Fig.3 by the potential difference at $x=0$ between
the thin drawn lines, denoted $\mu^{\rm av}_{\rm DMS}$ and
$\mu^{\rm av}_{\rm NMS}$, that extrapolate the voltage drop in the
bulk of the material towards the interface.

While in the NMS the splitting of the Fermi levels is symmetrical
because the conductivities for spin up and spin down electrons are
identical, in the DMS, the splitting for the majority-
($c^\uparrow \equiv \mu^\uparrow_{\rm DMS}(0)- \mu^{\rm av}_{\rm
DMS}(0)$) and the minority- ($c^\downarrow \equiv \mu^{\rm
av}_{\rm DMS}(0) - \mu^\downarrow_{\rm DMS}(0) $) spin electrons
can, in a one-dimensional model, be expressed as\cite{LMR,SST}

\begin{equation}
c^\uparrow , c^\downarrow = -\frac{\lambda_{\rm NMS}}{\sigma_{\rm
NMS}}\frac{I\beta(\beta \pm 1)}{(1+e^{-\frac{x_0}{\lambda_{\rm
NMS}}}+2\frac{\lambda_{\rm NMS}}{x_0} e^{-\frac{x_0}{\lambda_{\rm
NMS}}})+\frac{\lambda_{\rm NMS}\sigma_{\rm DMS }}{\sigma_{\rm
NMS}\lambda_{\rm DMS}}(1-\beta^2)}
\end{equation}

where in the numerator the plus(minus) sign applies to
$c^\downarrow$ ($c^\uparrow$), respectively. In Eq. 1,
$\lambda_{\rm DMS}, \lambda_{\rm NMS}, \sigma_{\rm DMS},
\sigma_{\rm NMS},$ are the spin flip length and the conductivity
in the DMS and the NMS respectively, $x_0$ is the spacing between
the contacts, $I$ is the current and $\beta$ is the degree of spin
polarization in the bulk of the contacts.  Note that $c^\uparrow$
and $c^\downarrow$ are defined setting $\mu^{\rm av}_{\rm DMS}(0)$
as reference level for the energy scale, i.e. $\mu^{\rm av}_{\rm
DMS}(0) = 0$. For the potential drop $\Delta U$ at the interface,
and the resulting magnetoresistance we simply have:

\begin{subequations}
\begin{equation}
\Delta U = \mu^{\rm av}_{\rm DMS}(0) - \mu^{\rm av}_{\rm NMS}(0) =
( c^\uparrow + c^\downarrow ) / 2,
\end{equation}
\begin{equation}
\Delta R = \Delta U / I.
\end{equation}
\end{subequations}

Eqs. 1 and 2 are quite general and describe spin injection in
metals as well as semiconductors - but only in the linear regime.
The magnitude of the Fermi-level splitting (and thus of $\Delta
U$) is different for different types of junctions: since the
spin-polarized current is driven solely by the spin accumulation,
the Fermi-level splitting has to be of the order of the current
imbalance between the spin channels times the resistivity of the
normal metal. When the magnet and the non magnet are both metals,
as, e.g., in CPP GMR\cite{cpp}, the splitting remains small even
for high currents due to the high conductivity. When the magnet is
metallic and the non-magnet is a semiconductor, the splitting is
determined by the material with the highest conductivity, and thus
is also small. This effect leads to negligible levels of spin
injection \cite{gs}, and is also known as the conductance
mismatch. When magnet and non magnet are both semiconductors as in
the large magnetoresistance effect\cite{LMR} discussed here, the
splitting can easily be in the range of mV.

In a magnetic metal the Fermi energy is typically of the order of
a few eV. A Fermi-level splitting of a few $\mu$V at the interface
will thus not influence the properties of the magnet, and one
expect the linear regime to hold up to quite high current
densities. In a DMS the situation is completely different. The
Fermi energy, the Zeeman splitting, and the Fermi level splitting
are in all the range of mV, and non-linear effects are to be
expected when bias voltages of similar magnitude are applied.

\begin{figure}
\centerline {\psfig{file=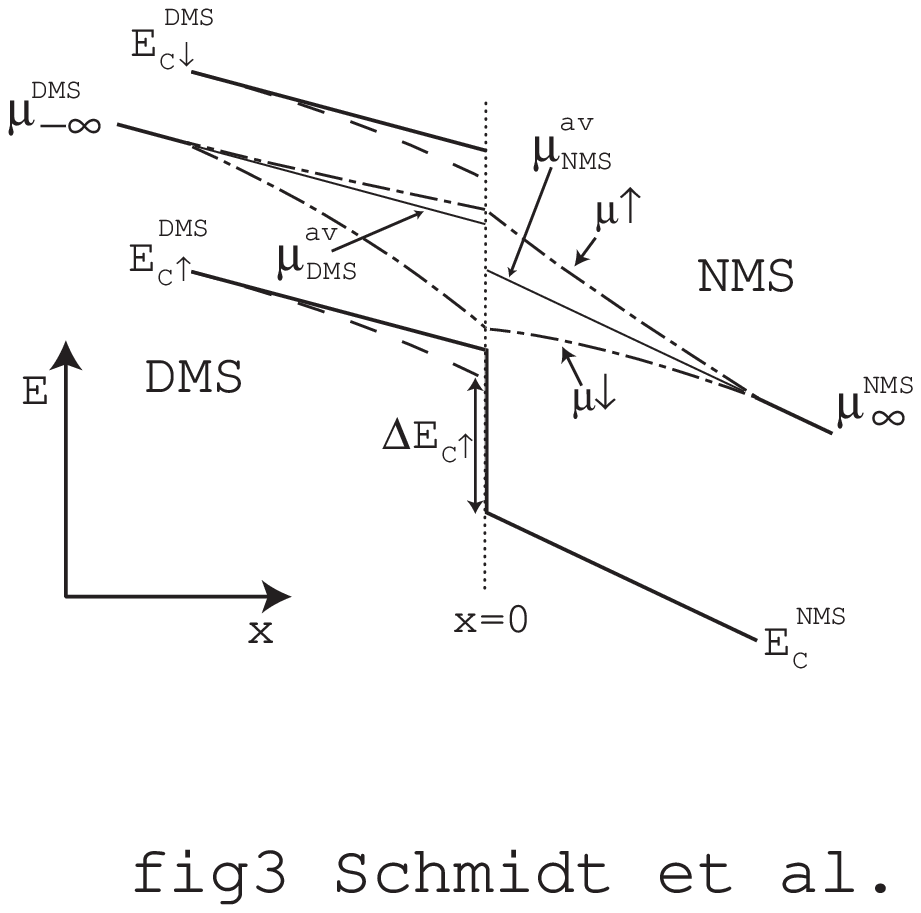,width=14.0cm}} \caption{Diagram
of the band bending at the spin-injecting DMS/NMS interface.
$\Delta E_{\rm C^\uparrow}$ denotes the location of the conduction
band offset between $E^{\rm DMS}_{\rm C^\uparrow}$ and $E^{\rm
NMS}_{\rm C}$ when bend-bending is taken in to account; all other
symbols are discussed in the text. Note the discontinuity between
$\mu^{\rm av}_{\rm DMS}$ and $\mu^{\rm av}_{\rm NMS}$ at the
junction ($x=0$), which is the potential difference $\Delta U$ in
Eqs. 2.}\label{interface}
\end{figure}

This situation, applied to the present experiment, is pictured in
some detail in Fig. \ref{interface}. The conduction band of the
NMS is some tens of mV below the conduction band of the DMS. The
conduction band level of the DMS is split by the Zeeman splitting
into two subbands,  $E^{\rm DMS}_{\rm C^\downarrow}$ and $E^{\rm
DMS}_{\rm C^\uparrow}$. From previous spin injection
experiments\cite{fiederling,LMR} and from spin flip Raman
scattering we know that the DMS is fully spin polarized at low
temperatures and moderate magnetic fields, which recent band
structure calculations understand as resulting from the formation
of an impurity band.\cite{Sato} This implies that the Fermi energy
is situated above the lower and at least a few mV below the upper
Zeeman level.
%We define the energy difference between the upper
%Zeeman level and the Fermi energy as $\Delta E^\downarrow$

As discussed above, spin injection will lead to the occurrence of
a 'built-in potential' $\Delta U$ at the interface. This is an
actual electrochemical potential step (i.e., not spin-dependent).
In order to preserve both charge conservation and the band offset
at the junction, $\Delta U$ has to be compensated by band bending
and charging at the interface. In Fig. 3, this is indicated by the
dashed lines emanating from $E^{\rm DMS}_{\rm C^\downarrow}$ and
$E^{\rm DMS}_{\rm C^\uparrow}$. (In principle, one also expects
band bending at the NMS side of the junction. For clarity, we have
not included this in Fig. 3, nor in the modelling we describe
below. Its inclusion is straightforward.)

Thus modified, Fig. 3 satisfies all boundary conditions imposed by
the band structure and the continuity equations for the
spin-polarized currents at the interface and for the total current
throughout the device. However, it is now obvious that the
properties of the DMS can be seriously affected by the band
bending. At the interface, the majority spin electrochemical
potential $\mu^\uparrow$ now approaches the upper Zeeman level
$E^{\rm DMS}_{\rm C^\downarrow}$. From the strong temperature
dependence observed for the LMR effect in Ref. \cite{LMR} we know
that this energy gap is small enough to strongly influence the
population of the upper Zeeman level and thus the bulk spin
polarization $\beta$ in the DMS, even at low temperatures. $\beta$
being close to one, however, is a prime prerequisite for injecting
a highly spin polarized current into the NMS. We can thus expect
the spin polarization injection (and thus the LMR effect) to
collapse as soon as the band bending starts to reduce $\beta$.

Since $\beta$ and $\Delta U$ depend on each other, a modelling of
the phenomena as a function of applied bias voltage should be done
in a self-consistent manner. However, one can formulate a
mathematically consistent description when starting from a given
value of $\Delta R / R$ in the linear response regime (this is the
main reason for presenting the data  emanating from the same
$\Delta R/R$ value in Fig. 2.) We first use Eqs. 1 and 2 to
calculate the bulk polarization $\beta$ in the DMS. In the linear
regime, the bulk value of $\beta$ equals $\beta (x=0)$, the spin
polarization at the interface. Assuming Boltzmann statistics, we
then directly have the energy splitting between $E^{\rm DMS}_{\rm
C^\downarrow}(0)$ and $\mu^\uparrow(0)$ from

\begin{equation}
\beta (x=0) = \tanh ((E^{\rm DMS}_{\rm C^\downarrow}(0)- \Delta U
- \mu^{\uparrow}(0))/2k_{\rm B}T),
\end{equation}

where of course $\Delta U = 0$ for infinitesimally small bias. For
modelling the dependence on $V_{\rm bias}$, we gradually increase
$\Delta U$ (note that at this point we assume all band-bending to
occur in the DMS), calculate the reduced $\beta (x=0)$ using Eq.
3, and substitute this value for the bulk polarization in Eqs. 1
and 2 to calculate $\Delta R/ R$. At the same time, $\Delta U$ can
be converted in a voltage drop across the junction, $V_j$.

The latter quantity is conveniently accessible for comparison with
the experiment. This is because  $\Delta R \rightarrow 2 \lambda
_{\rm NMS} / \sigma_{\rm NMS}$ for $ B, ( x_0/\lambda_{\rm NMS} )
\rightarrow \infty $, as can easily be verified from Eqs. 1 and 2.
Experimentally, we have (within our  1-dimensional modelling)
$\sigma_{\rm NMS} = 2.5 \times 10^{-4} \Omega^{-1}$cm, yielding
$\lambda_{\rm NMS} = 1.25 \mu$m. For both the experimental (Fig.
2a) and theoretical (Fig. 2b)  plots of the non-linear behavior,
we may now calibrate the voltage axis according to $V_j = I \Delta
R + V_{\rm bias}(\lambda_{\rm NMS}/x_0)$.

As to the remaining parameters in our model, we have from
experiments on single DMS layers that $\sigma_{\rm DMS}$,
converted to the 1-dimensional model equals $ 1.0 \times 10^{-4}
\Omega^{-1}$cm. The only free parameter now left in the model is
$\lambda_{\rm DMS}$. Since there is no easy method to measure
$\lambda_{\rm DMS}$, and moreover its magnetic field dependence is
unknown, we simply have set the ratio $\lambda_{\rm
NMS}\sigma_{\rm DMS } / \sigma_{\rm NMS}\lambda_{\rm DMS} $ in
Eqs. 1 equal to 1, yielding $\lambda_{\rm DMS} = 0.5 \mu$m.

The modelling of the band bending effect as described above leads
to the plots shown in Fig. 2b. We find that indeed a few mV of
voltage drop across the junction are enough to reduce the spin
polarization of the injected current to almost zero. The computed
curves closely resemble the experimental results in shape,
magnitude, voltage range, and temperature dependence, and confirm
the validity of the modelling discussed above.

At this point, we should address the drift effects introduced by
Yu and Flatt\'{e}\cite{flatte}, that also can induce a reduction
of $\Delta R / R$ in our experiments. For the highly (i.e., above
the metal-insulator transition) doped samples used here, it is
easy to show that drift effects can only be expected for much
higher electric fields than those needed to actually observe the
non-linearities. Moreover, within the drift model one would not
expect any temperature dependence for degenerate semiconductors,
again in contradiction with the experiments. We do, however, have
some indications that drift effects actually occur in certain
ranges of operation. Note, e.g., that $\Delta R / R $ is not
completely suppressed for high $V_j$ in the modelling curves of
Fig. 2b, while experimentally $\Delta R / R $ vanishes entirely.
The drift mechanism can actually explain this behavior, be it for
fields that are an order of magnitude larger than those needed to
induce the initial non-linearities.

In conclusion, we have shown that when spin injection into
semiconductors is used beyond the regime of linear response, the
splitting of the electrochemical potentials can influence the band
bending in spin-injecting junction. For finite 'built-in
potentials' $\Delta U$ it is important that the energy difference
$( E^{\rm DMS}_{\rm C^\downarrow} - \mu^\uparrow (0))$ is kept
sufficiently large to ensure large spin polarization at the
interface. This observation may seriously limit the operational
voltage of spintronic devices and forbid applications in power
electronics. Appropriate tailoring of the band structure may be
possible to circumvent the problems described here.

\acknowledgments

We acknowledge the financial support of the Bundesministerium
f\"ur Bildung und Forschung, the DFG (SFB 410) and the DARPA SPINS
program. We would like to thank V. Hock for device fabrication,
and M. Flatt\'{e}, K. Flensberg, and E. Rashba for useful
discussions.

\newpage

\begin{thebibliography}{99}
\bibitem{4probe}Since the spin splitting is maximum at the interface
and reduces appreciably within the NMS, the actual geometry used
by us is necessary in order to measure the full spin-induced
boundary resistance.
\bibitem{flatte2}For larger Fermi-level splitting or higher
electric field, this statement is no longer correct, and the
electron transport must be modelled using a drift model as
discussed in Ref. \protect\onlinecite{flatte}.
\end{thebibliography}
\end{document}